# Offset Dual Microcombs in a Single Microresonator: Kerr–Raman Interleaved Combs

Renhong Gao[1], Qifeng Hou[2], Xinzhi Zheng[1], Bin Li[3], Guanghui Zhao[2], Yingnuo Qiu[2], Xinke Xing[2], Boyang Nan[1], Yixuan Yang[2], Saisai Sun[2], Kunpeng Jia[3,§], Zhenda Xie[3,*], Jintian Lin[2,‡], Shining Zhu[3], and Ya Cheng[1,4,5,6,7,8,†]

*[1]The Extreme Optoelectromechanics Laboratory (XXL), School of Physics, East China Normal University, Shanghai 200241, China*

*[2]State Key Laboratory of Ultra-intense Laser Science and Technology, Shanghai Institute of Optics and Fine Mechanics, Chinese Academy of Sciences, Shanghai 201800, China*

*[3]National Laboratory of Solid State Microstructures, School of Electronic Science and Engineering, College of Engineering and Applied Sciences, School of Physics, Research Institute of Superconductor Electronics (RISE) & Key Laboratory of Optoelectronic Devices and Systems with Extreme Performances of MOE, Key Laboratory of Intelligent Optical Sensing and Manipulation, Ministry of Education, and Collaborative Innovation Center of Advanced Microstructures, Nanjing University, Nanjing 210093, China*

*[4]State Key Laboratory of Precision Spectroscopy, East China Normal University, Shanghai 200062, China*

*[5]Shanghai Research Center for Quantum Sciences, Shanghai 201315, China*

*[6]Hefei National Laboratory, Hefei 230088, China*

*[7]Collaborative Innovation Center of Extreme Optics, Shanxi University, Taiyuan 030006, China*

*[8]Collaborative Innovation Center of Light Manipulations and Applications, Shandong Normal University, Jinan 250358, China*

*[§]E-mail: jiakunpeng@nju.edu.cn*

**E-mail: xiezhenda@nju.edu.cn*

*[‡]E-mail: jintianlin@siom.ac.cn*

*[†]E-mail: ya.cheng@siom.ac.cn*

# Abstract

Integrated Kerr microcombs have emerged as a central platform for chip-scale nonlinear photonics. One of their most compelling applications is dual-comb technology, which enables fast, high-resolution, coherent optical spectroscopy, precision metrology, ranging, and ultrasensitive sensing by interfering two frequency combs with slightly different repetition rates. However, no effective strategy yet exists to synergistically excite multiple spatial mode families, manage mode competition and energy partitioning, and precisely engineer dispersion, leaving offset co-propagating dual-microcombs in a single resonator elusive. Here we demonstrate a new mechanism for offset dual microcombs formation via Kerr–Raman interleaving for the first time, by leveraging single Raman-active thin-film lithium niobate microrings. Two spatial mode families with high loaded Q factors up to $5\times10^6$ are dispersion engineered to have nearly identical FSRs but a designed relative frequency offset. Stimulated Raman scattering efficiently generates a Raman comb in one family alongside a Kerr comb in the other. The two combs interleave with a fractional offset to form a broadband, densely interleaved spectrum. Pumping the microring at only 320 mW yields Kerr–Raman interleaved dual combs both featuring nearly identical repetition rates around 102 GHz with a small difference of 675 MHz, a relatively flat envelope exceeding 29 nm and an intrinsic comb-line linewidth of 410 Hz. Our results establish offset dual-comb generation in a single microresonator, enabling compact dual-comb sources for chip-scale precision spectroscopy, Vernier-enhanced sensing, and advanced quantum photonics.

Integrated Kerr microcombs—characterized by small footprint, low power consumption, high robustness, high conversion efficiency, and scalability—have become central to chip-scale nonlinear photonics [1,2]. One of their most compelling applications is dual-comb technology, which enables fast, high-resolution, coherent optical spectroscopy, precision metrology, ranging, and ultrasensitive sensing by interfering two frequency combs with slightly different repetition rates [3-5]. However, practical on-chip dual-comb sources remain challenging to realize.

One strategy has been proposed to generate on-chip dual-comb systems generally relied on two separate combs, often requiring two distinct microresonators and complex control systems for pump and propagation management [6-12]. Despite these rapid advances, these approaches result in severely increased cost, power consumption, environmental instability, and system complexity. A more elegant approach is to generate two independent combs within the same microresonator under a single continuous-wave pump laser, which would significantly enable low-threshold operation, improved robustness, and Vernier-enhanced sensing. Such an approach, however, remains fundamentally difficult because no effective strategy currently exists to synergistically excite multiple spatial mode families, manage mode competition and energy partitioning, and precisely engineer dispersion. Consequently, offset co-propagating dual-microcombs in a single resonator have remained elusive.

Here, we demonstrate a new mechanism for offset dual-microcomb formation in a single Raman-active thin-film lithium niobate (TFLN) microring resonator. Two spatial mode families with high loaded Q factors up to $5\times10^6$ are dispersion engineered to exhibit nearly identical free-spectral ranges (FSRs) but a designed relative frequency offset. By leveraging stimulated Raman scattering, a Raman comb is excited in one family alongside a Kerr comb in the other; the two combs interleave with a fractional offset to form a densely interleaved spectrum. Pumping at 320 mW yields Kerr–Raman interleaved dual combs that exhibit nearly identical repetition rates around 102 GHz, a small difference of 675 MHz, a flat envelope exceeding 29 nm in the telecom band, and an intrinsic comb-line linewidth of 410 Hz. This work establishes a robust pathway for on-chip co-propagating dual-comb generation in single microresonators, significantly advancing dual-comb systems toward simplified architectures, enhanced robustness, and scalable integration.

## Results

### Design and fabrication of the single microresonator

Monolithic TFLN microrings were designed with tailored dispersion profiles to enable simultaneous dual-comb generation with a slight repetition-rate difference. The dispersion engineered microrings were fabricated with high Q factors in a Z-cut TFLN wafer using femtosecond laser photolithography assisted chemo-mechanical etching (PLACE) [13,14]. The scanning-electron-microscope (SEM) image of the fabricated monolithic microring is depicted in the left inset of Fig. 1, where the microring is side-coupled to a bus waveguide for optical coupling. The microring possesses a diameter of 400 μm, a top width of 2 μm, and a wedge angle of 9.27°. Because the produced sidewalls are ultra-smooth, with a typical average surface roughness of ~0.2 nm, this microring supports multiple high-Q spatial mode families in the telecom band. For example, the loaded Q factor of transverse-electric (TE) modes in the telecom band was measured to be $5.1\times10^6$ (right inset of Fig. 1). Such high Q factors enable strong intracavity field builtup, thereby facilitating both low-threshold SRS and efficient Kerr nonlinear processes.

## Experimental setup

The experimental setup for the dual-comb generation is schematically illustrated in Fig. 1. A tunable continuous-wave (CW) laser operating in the telecom band served as the pump source. After passing through a polarization controller (PC), the pump light was amplified by an erbium-doped fiber amplifier (EDFA), filtered by a narrow-bandwidth band-pass tunable filter, and coupled into the TFLN microring resonator through the bus waveguide using a lensed-fiber coupling system. During the experiment, the pump wavelength, polarization state, and coupling condition were carefully optimized to ensure efficient excitation of the target mode families. The output from the microring was divided into two paths using a 90:10 fiber splitter. One path was directed to an optical spectrum analyzer (OSA) for spectral characterization, while the other was detected by a high-speed photodetector (PD) and analyzed with an electrical spectrum analyzer (ESA) to characterize the radio-frequency (RF) beat-note signals of the generated combs. To cover the full spectral ranges of the generated dual-combs, two optical spectral analyzers (OSAs) were employed, with OSA1 covering 600–1700 nm and OSA2 covering 1200–2400 nm. The spectra recorded by the two instruments were subsequently stitched together using their overlapping wavelength ranges.

## Dual-comb generation

When the pump laser was tuned to 1551.28 nm with a pump level of 320 mW, two independent optical frequency combs were simultaneously observed, as confirmed by the detected spectrum in Fig. 2(a). In the spectrum, one comb whose envelope is marked by the red curve, spans from ~1329 nm to ~1664 nm. Specifically, this comb aligns with the pump light, with its comb lines symmetrically distributed around with the pump light, with a wavelength spacing of approximately 0.82 nm between adjacent comb lines, as shown in Fig. 2(b). While the other comb, whose envelope is marked by the purple curve, spans a broader spectral range, from 1300 nm to 1954 nm. The corresponding wavelength spacing between adjacent comb lines near the pump wavelength is also approximately 0.82 nm, indicating that these two combs possess similar repetition rates. However, the comb is not aligned with the pump light; instead, it is offset by 0.278 nm, as verified in Fig. 2(c). In other words, the two combs are offset from each other and do not merge spectrally. Moreover, both combs exhibit relatively flat spectral profiles in the telecom band, as shown in Fig. 2(b). Specifically, the combs possess relatively flat spectra exceeding 29 nm, spanning from 1531.56 nm to 1560.8 nm. The overall optical conversion efficiency of the two combs reaches approximately 13.8%. Therefore, these two combs simultaneously exhibit flat spectral profiles and approximately equal repetition rates, both of which are crucial for dual-comb technology applications.

Figure 2(d) shows a low-frequency RF beat-note microwave spectrum of the generated combs, which exhibits no pronounced direct-current (DC) noise within the measurement range. To access the coherence properties of the generated combs, a single comb line was selected from the output spectrum, isolated using a narrow-band optical filter and then used as the input signal for frequency-noise characterization. The frequency-noise power spectral density (PSD) of the selected comb line was measured as a function of Fourier frequency (Fig. 2(e)). The measured PSD gradually decreases with increasing Fourier frequency and approaches a constant background level (i.e., white noise level) at high frequency offset. By integrating the measured frequency-noise spectrum, the effective linewidth of the selected comb line was extracted to be 410.63 Hz. These results indicate that the dual-comb generation preserves high coherence.

**Dynamic evolution of the dual-comb generation varied with incremental pump power**

To elucidate the formation dynamics of dual-comb generation in the single microring resonator, we investigated the spectral evolution with the gradually increasing pump power. Throughout the measurements, the pump wavelength, coupling condition, and polarization state were kept constant, while the output spectra were systematically recorded at different pump powers. As the pump power increases, SRS signals first emerge, followed by the formation and spectral expansion of two distinct optical frequency combs. Figure 3(a) shows the spectrum obtained at a pump power of ~55 mW, revealing the emergence of two nonlinear signals at 1610.12 nm and 1511.68 nm, respectively. These two signals should be attributed to Stokes and anti-Stokes SRS [13,15,16], involving two E-symmetry optical phonon branches of lithium niobate at approximately 237 $cm^{-1}$ and 169 $cm^{-1}$, respectively.

As the pump power increased to beyond 75 mW, Kerr parametric sidebands began to emerge independently around these two SRS signals and the pump light, leading to the generation of the Raman assisted Kerr combs (i.e., Raman combs) and the pure Kerr comb, as shown in Fig. 3(b). These sidebands broaden further with increasing pump power, as evidenced in Fig. 3(c). When the pump power reaches 130 mW and even higher, the sidebands triggered by the Stokes and anti-Stokes SRS signals begin to merge, covering a range from approximately 1475 nm to 1650 nm, which corresponds to Raman comb generation, as shown in Figs. 3(c) and (d). In addition, other comb lines appear around 1425 nm and 1700 nm, facilitating the spectral broadening of the Raman comb. While the pure Kerr comb directly triggered by the pump light broadens to a spectral range of 1525 nm to 1575 nm. When the pump power reaches 290 mW, the Raman comb extends from 1350 to 1700 nm, whereas the coverage of the pure Kerr comb ranges from 1370 nm to 1600 nm, as depicted in Fig. 3(e).

The dynamic spectral evolution with varying pump power clearly reveals that the dual-comb generation arises from two correlated but physically distinct nonlinear processes. The pump field directly drives Kerr effect, producing the pure Kerr comb from the pump mode family. Meanwhile, the pump light efficiently excites the Stokes and anti-Stokes SRS processes, which act as secondary pump sources to initiate Kerr nonlinear process in another spatial mode family, thereby generating of a Raman comb that is spectrally distinct from the pure Kerr comb.

**Dispersion analysis and numerical simulation**

To further reveal the physical mechanism underlying the dual-comb generation in the single microring, we analyzed the dispersion profiles of the cavity modes involved in the nonlinear frequency conversion processes. Transmission spectrum of the cavity modes was characterized by scanning the tunable laser, and the wavelength was calibrated using both an unbalanced Mach-Zehnder interferometer and an HCN wavelength reference. After wavelength calibration, four spatial mode families, including $TE_0$, $TE_1$, $TM_1$ (i.e., first-order transverse-magnetic mode), and $TM_2$, were identified from the resonance spectra. Among them, the $TE_0$ and $TE_1$ mode families exhibit average FSRs of 0.81996 nm and 0.81454 nm, respectively, corresponding to repetition rates of approximately 102 GHz with a slight difference of ~675 MHz. The eigenmodes of different spatial mode families were calculated using the finite-element method (FEM), and their corresponding integrated dispersion profiles were then extracted. Based on the experimentally extracted resonance wavelengths and FEM simulated calculations, the pump mode was attributed to a high-order TE mode of $TE_{1,1652}$, where the subscripts (1,1652) denote the radial and azimuth mode numbers, respectively. The Stokes and anti-Stokes modes were resonant with the fundamental TE modes of $TE_{0,1602}$ and $TE_{0,1721}$, respectively. The calculated integrated dispersion curves and mode profiles of the pump-related $TE_1$ mode family and Raman-related $TE_0$ mode family are presented in Figs. 4(a) and (b), respectively. In the telecom band, the $TE_1$ mode family exhibits anomalous dispersion, providing favorable phase-matching conditions for the pure Kerr comb generation observed experimentally. In contrast, the $TE_0$ mode family exhibits relatively flat normal dispersion for facilitating broadband Raman comb generation. Despite the substantially different integrated dispersion profiles of the pump and SRS modes, these modes can be efficiently bridged through the excitation of the SRS process in the dispersion-engineered microring. Consequently, these two distinct spatial mode families within the single microring resonator can support the formation of separate combs, collectively enabling dual-comb generation.

We further simulated the output spectra through solving the Lugiato–Lefever equation (LLE) [17], using the calculated mode dispersion profiles and pump conditions. The simulated spectra in Figs. 4(e)-(g) reproduce the major spectral features observed experimentally across different wavelength regions. The numerical reproduction of the Raman comb is depicted in Figs. 4(e) and (f). The simulated result agrees well the experimentally observed envelope, confirming the effective broadband spectral expansion. Therefore, the generated SRS fields act as effective intracavity pump sources for subsequent broadband Raman comb generation. Meanwhile, the simulated spectrum of the pure Kerr comb is plotted Fig. 4(g), which also well reproduces the main spectral characteristics observed experimentally. This agreement demonstrates that the high-order spatial modes with anomalous dispersion can be harnessed to generate Kerr comb. The remaining small discrepancies between simulation and experiment mainly arise from the simplified numerical model, which does not fully incorporate experimental factors such as pump detuning, coupling conditions, thermal effects, and transient nonlinear dynamics.

**Discussion and conclusion**

It is worth noting that, Raman combs have been extensively demonstrated in various photonic platforms [18,19]. In these experimental demonstrations, however, the pump mode and the excited SRS mode usually belong to the same mode family, thereby restricting the dual-comb generation. Moreover, dual-comb generation involving two spatial mode families requires not only that both families possess suitable dispersion profiles but also that they be efficiently bridged by the SRS process so that energy and momentum conservation are satisfied. Consequently, dispersion engineering is challenging, and dual-comb generation in a single microresonator under single-laser pumping has not yet been demonstration. In contrast to conventional dual-comb architectures, our approach requires only one microring resonator driven by a single unidirectional CW pump, thereby eliminating the need for synchronization between independent resonators and for complex external feedback, routing, and mixing control systems. More importantly, it actively exploits the intrinsic giant Raman nonlinearity of TFLN. Rather than acting as a competing mechanism that suppresses Kerr comb formation, the Raman process is transformed into an enabling mechanism for generating a second optical frequency comb [20,21]. This strategy provides a practical route toward highly integrated dual-comb sources based on a single microresonator.

In conclusion, dual-comb generation is reported in a single dispersion engineered TFLN microring. Two distinct spatial cavity mode families with appropriate dispersion profiles were leveraged and bridged through SRS process. By pumping the microring with a single CW laser, dual-comb generation with similar repletion-rates of approximately102 GHz is achieved through the pure Kerr effect and Raman assisted Kerr comb generation. Notably, TFLN is not only a highly Raman-active platform but also combines tight optical confinement with outstanding optical properties, including high quadratic nonlinearity [22-26], a strong electro-optic coefficient [26,27], and strong acousto-optic/piezoelectric effects [28,29], making it an emerging photonic platform [30]. Further exploration of these properties of TFLN can substantially enhance the performance and functionality of dual-comb generation, such as multi-chromatic outputs, high-speed modulating, electro-optic reconfiguration, and scalable integration, without requiring complex heterogeneous integration.

We thank Prof. Chen Yang from Nankai University for providing the bandwidth-tunable optical filter, Prof. Hairun Guo from Shanghai University for providing the phase noise analyzer, and Prof. Xingchen Ji from Shanghai Jiao Tong University for helpful discussion.

**Captions of figures:**

FIG. 1 (Color online) Experimental setup for dual-comb generation in a unidirectionally pumped TFLN microring resonator. CW, continuous-wave; EDFA, erbium- doped fiber amplifier; CCD, charge-coupled device; OSA, optical spectrum analyzer; PD, photodetector; ESA, electrical spectrum analyzer. Left Inset: Scanning electron microscope (SEM) image of the fabricated microrings. Right Inset: Typical transmission spectrum of a TE mode in the telecom band, showing a loaded Q factor as high as $5.1\times10^6$.

FIG. 2 (Color online) Dual-comb generation. (a) Optical spectrum of the dual-combs. (b) Zoom-in spectrum range in the telecom band. (c) Enlarged spectrum range around the pump wavelength. (d) RF microwave spectra of the signal (Sig.) and the background (Bg.). (e) Phase noise of one comb line measured by short-delayed self-heterodyne methods.

FIG. 3 (Color online) Spectral evolution. (a) Realization of both the Stokes and anti-Stokes SRS signals for Raman comb generation. (b)-(e) Experimentally measured spectral evolution of the dual-comb generation under different pump powers.

FIG. 4 (Color online) Dispersion engineering and numerical analysis of dual-comb generation. (a)-(d) Numerically calculated (blue curve) and experimentally measured (red dots) integrated dispersion profiles. Insets in (b) and (d): Electric-field mode profiles of the $TE_1$ and $TE_0$ mode families, respectively. (e)-(g) Comparison between the simulated and experimentally measured spectra around the Stokes, anti-Stokes, and pump wavelength regions, respectively, based on the calculated mode dispersion profiles.

**Fig. 1**

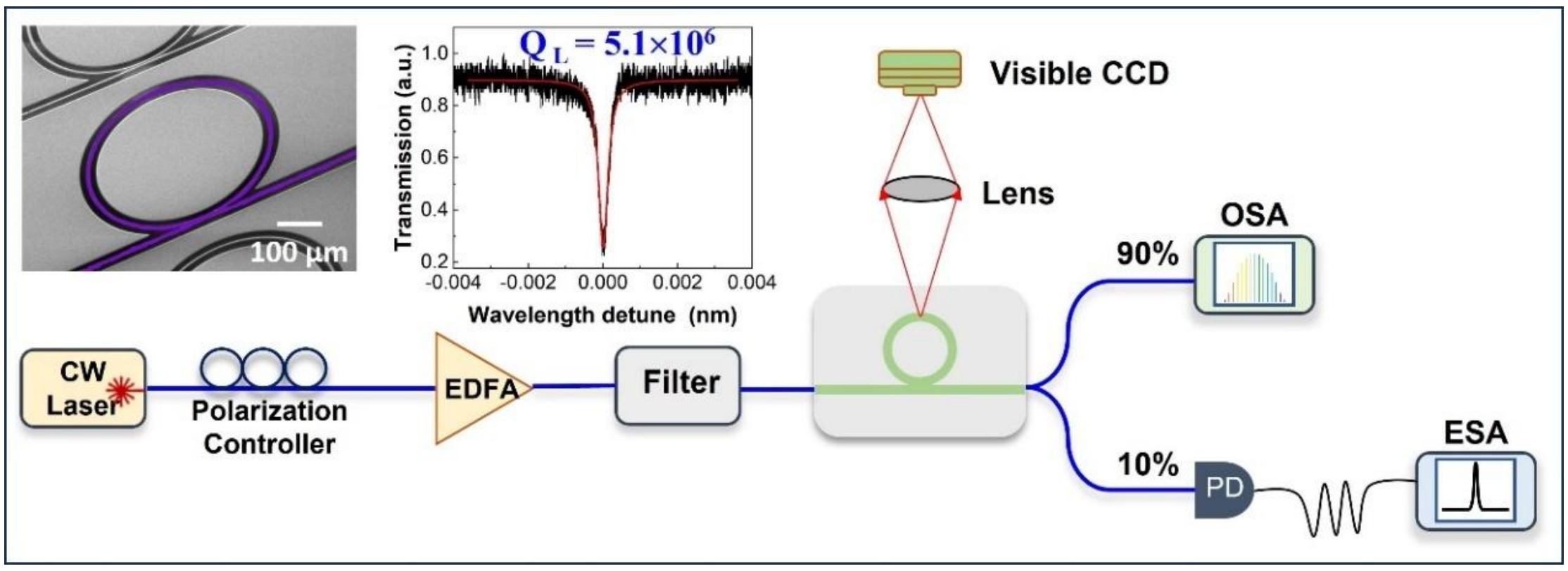

100 μm
$Q_L = 5.1\times10^6$
Transmission (a.u.)
1.0
0.8
0.6
0.4
0.2
-0.004
-0.002
0.000
0.002
0.004
Wavelength detune (nm)
Visible CCD
Lens
OSA
90%
CW Laser
Polarization Controller
EDFA
Filter
10%
PD
ESA

**Fig. 2**

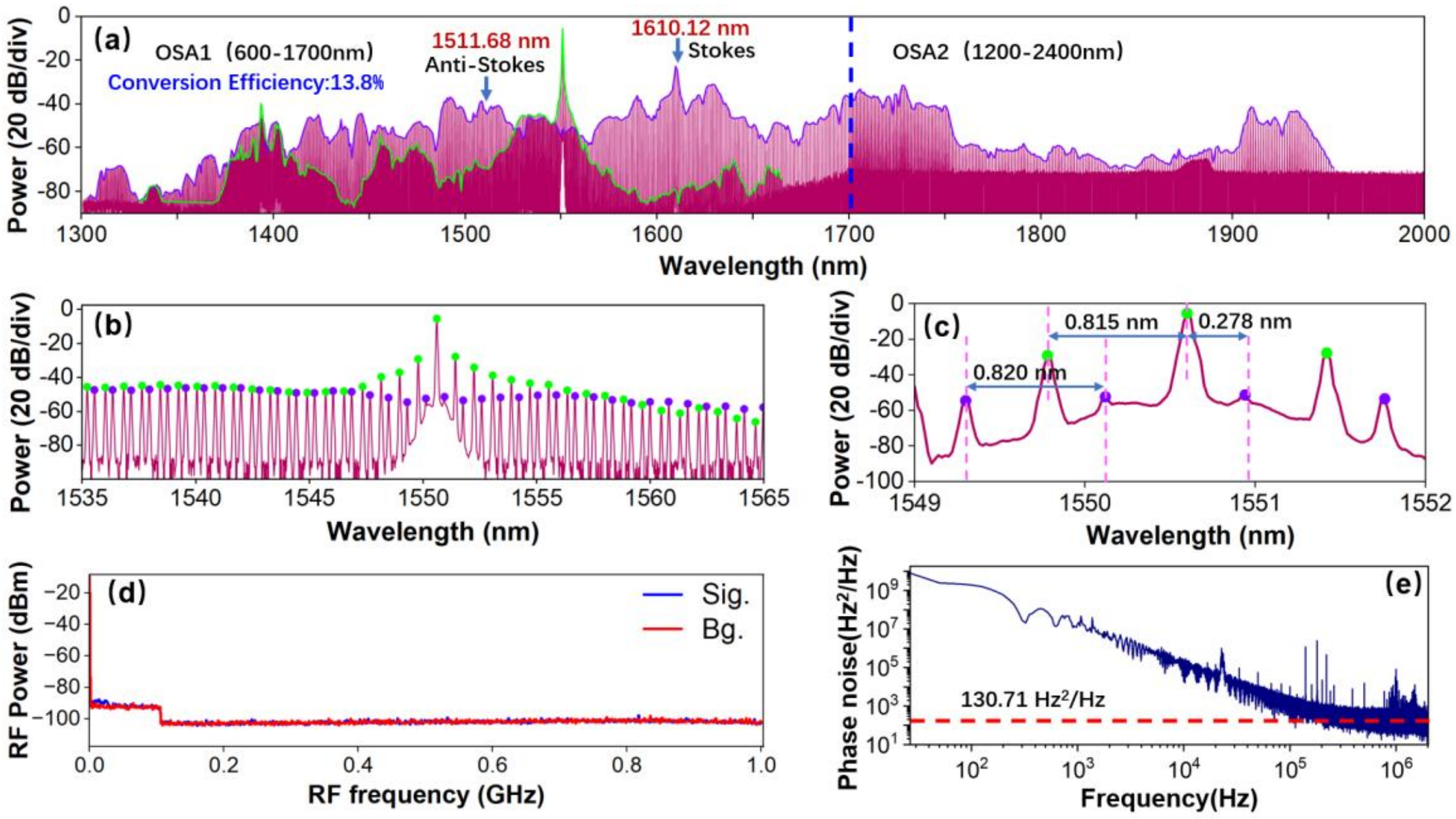
(a)
OSA1 (600-1700nm)
Conversion Efficiency:13.8%
1511.68 nm
Anti-Stokes
1610.12 nm
Stokes
OSA2 (1200-2400nm)
Power (20 dB/div)
Wavelength (nm)
(b)
Power (20 dB/div)
Wavelength (nm)
(c)
0.815 nm
0.278 nm
0.820 nm
Power (20 dB/div)
Wavelength (nm)
(d)
Sig.
Bg.
RF Power (dBm)
RF frequency (GHz)
(e)
130.71 Hz²/Hz
Phase noise(Hz²/Hz)
Frequency(Hz)

**Fig. 3**

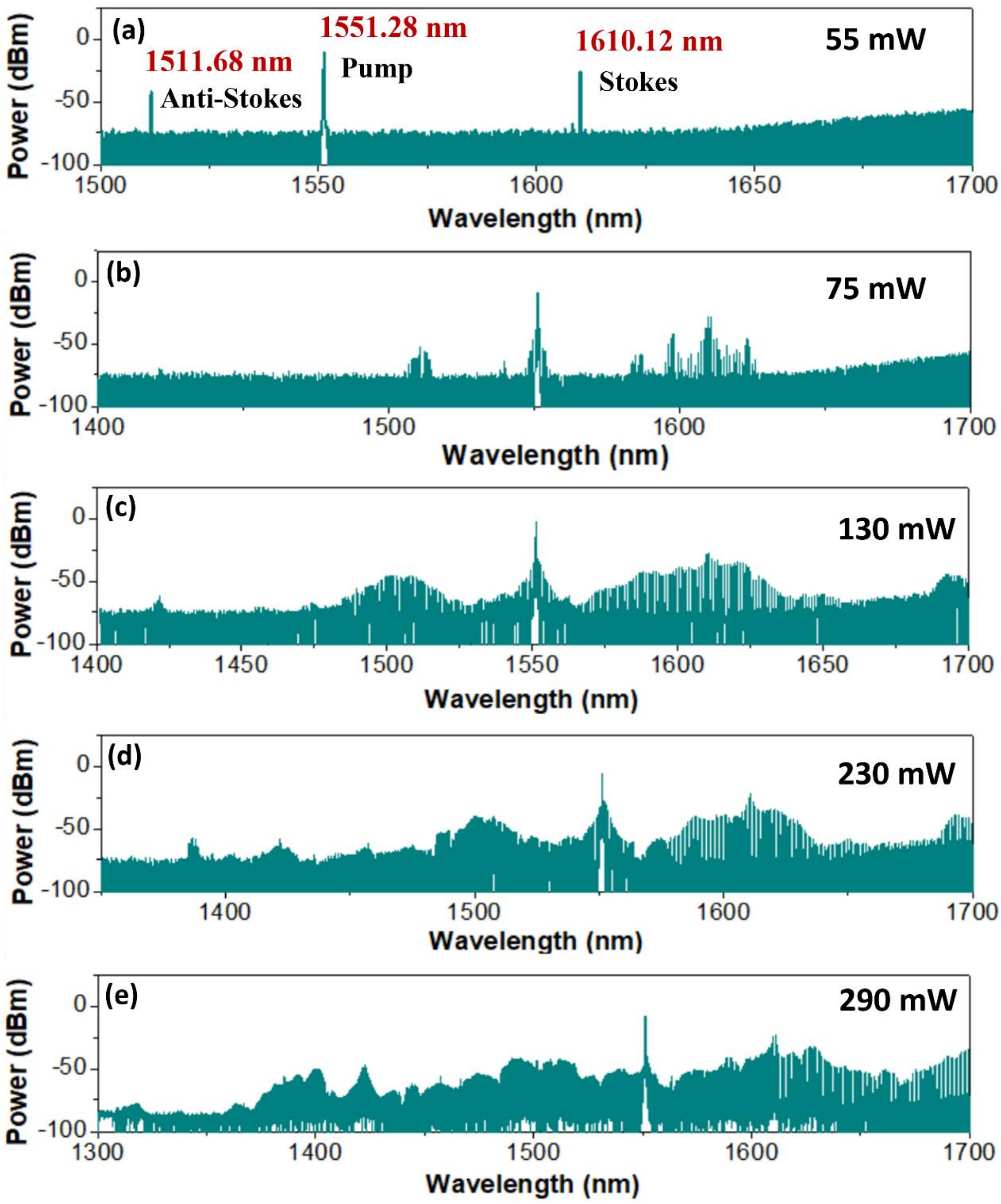
(a)
1511.68 nm
Anti-Stokes
1551.28 nm
Pump
1610.12 nm
Stokes
55 mW
Power (dBm)
Wavelength (nm)
(b)
75 mW
Power (dBm)
Wavelength (nm)
(c)
130 mW
Power (dBm)
Wavelength (nm)
(d)
230 mW
Power (dBm)
Wavelength (nm)
(e)
290 mW
Power (dBm)
Wavelength (nm)

**Fig. 4**

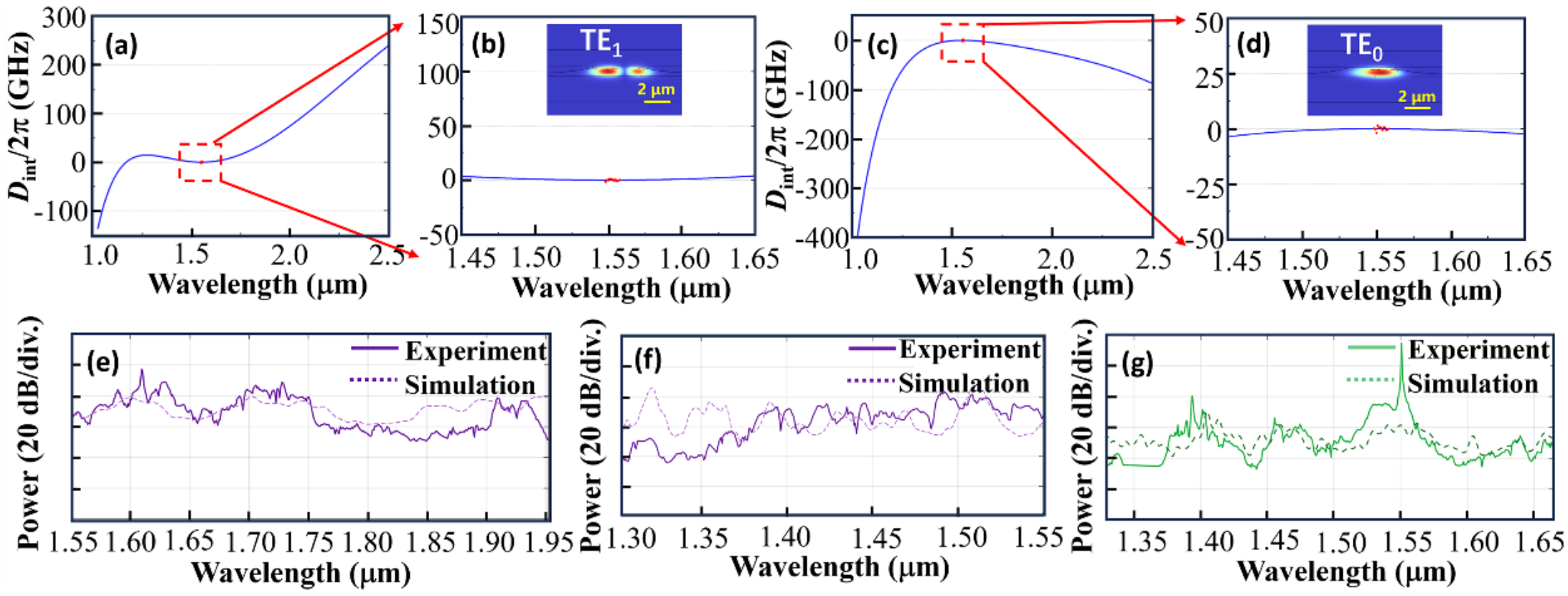

(a)
(b)
(c)
(d)
(e)
(f)
(g)
TE1
TE0
2 μm
Dint/2π (GHz)
Wavelength (μm)
Power (20 dB/div.)
Experiment
Simulation